\documentclass[pdflatex,sn-nature]{sn-jnl}

\usepackage{graphicx}%
\usepackage{multirow}%
\usepackage{amsmath,amssymb,amsfonts}%
\usepackage{amsthm}%
\usepackage{mathrsfs}%
\usepackage[title]{appendix}%
\usepackage{xcolor}%
\usepackage{textcomp}%
\usepackage{manyfoot}%
\usepackage{booktabs}%
\usepackage{algorithm}%
\usepackage{algorithmicx}%
\usepackage{algpseudocode}%
\usepackage{listings}%
\usepackage[switch]{lineno}

\theoremstyle{thmstyleone}%
\theoremstyle{thmstyletwo}%

\theoremstyle{thmstylethree}%

\newcommand\chandra{{\it Chandra~}}
\newcommand{\arcsec}{\ensuremath{^{\prime\prime}}}
\newcommand{\arcmin}{\ensuremath{^\prime}}

\usepackage{natbib}

\begin{document}

\title[Article Title]{A multi-wavelength study of nearby starburst galaxy M 82}


\author*[1]{\fnm{Nilkanth} \sur{Vagshette}}\email{nilkanth1384@gmail.com}  

\author*[2]{\fnm{Satish} \sur{Sonkamble}}\email{satish04apr@gmail.com}

\author[3]{\fnm{Madhav} \sur{Patil}}

\author[4]{\fnm{Sachindra} \sur{Naik}}

\author[2,5]{\fnm{Ilani} \sur{Loubser}}

\affil[1]{\orgdiv{Department of Physics and Electronics}, \orgname{Maharashtra Udayagiri Mahavidyalaya}, \orgaddress{\city{Udgir}, \postcode{413517}, \state{Maharashtra}, \country{India}}}

\affil[2]{\orgdiv{Centre for Space Research}, \orgname{North-West University}, \orgaddress{\city{Potchefstroom}, \postcode{2520}, \state{North West}, \country{South Africa}}}

\affil[3]{\orgdiv{School of Physical Sciences}, \orgname{Swami Ramanand Teerth Marathawada University}, \orgaddress{\street{Street}, \city{Nanded}, \postcode{431606}, \state{Maharashtra}, \country{India}}}

\affil[4]{\orgdiv{Astronomy \& Astrophysics Division}, \orgname{Physical Research Laboratory}, \orgaddress{\city{Ahmedabad}, \postcode{380009}, \state{Gujarat}, \country{India}}}

\affil[5]{\orgname{National Institute for Theoretical and Computational Sciences (NITheCS)}, \orgaddress{\city{Potchefstroom}, \postcode{2520}, \state{North West}, \country{South Africa}}}



\abstract{We present a multi-wavelength study of the nearby starburst galaxy M~82 by combining high-resolution Far-ultraviolet (FUV) imaging from the Ultra-Violet Imaging Telescope (UVIT) onboard AstroSat and archival \chandra X-ray observations. Using FUV flux measurements, we estimate a spatially-resolved star formation rate (SFR) across several star-forming clumps within a radius of $\sim$3.6 kpc, finding a total SFR of 0.022 M$_{\odot}$ yr$^{-1}$. The H$_{\alpha}$ recombination line flux yields an SFR of $\sim$0.010 M$_{\odot}$ yr$^{-1}$, while the infrared-based SFR derived from 24~$\mu\mathrm{m}$ emission is significantly higher at 16 - 18 M$_{\odot}$ yr$^{-1}$, suggesting that a substantial fraction of star formation in M 82 is heavily dust-obscured. Morphological comparison of FUV, H$_{\alpha}$, mid-infrared, and soft X-ray emission reveals a strong spatial correlation, tracing multi-phase outflows along the galaxy’s minor axis. X-ray spectral analysis using a three-temperature \texttt{VAPEC} model shows enhanced abundances of Ne, Mg, Si, and S, consistent with enrichment from Type-II supernovae. These results demonstrate the importance of combining UV, optical, IR, and X-ray observations to probe both obscured and unobscured star formation, the metal enrichment, and the outflow-driven evolution of starburst galaxies.}




\maketitle

\section{Introduction}\label{sec1}

The formation and evolution of galaxies is one of the most interesting phenomena in the universe, and the measurement of Star Formation Rate (SFR) provides important evidence for the investigation of the evolution of galaxies. The merging and interaction among galaxies provides intense starbursts in the cores of galaxies. Starburst galaxies typically have high star formation rates, and the measurement probes hot OB-type stars \citep{2019A&A...631A..51P}. These youngest stellar populations in galaxies emit most of their energy in the Ultra-Violet (UV) band \citep{2012ARA&A..50..531K}. Therefore, imaging studies of such galaxies using high spatial resolution telescopes in far-ultra-violet (FUV) and near-ultra-violet (NUV) helps us to locate the active star-forming regions in galaxies. The UV radiation from hot OB stars, short-ward of the Lyman limit, ionised the surrounding gas to produce ionized hydrogen HII regions, and this ionized gas recombines and produces emission lines. As such, this recombination emission line flux provides an estimate of ionisation radiation in the H$_{\alpha}$ domain, which is commonly used to estimate the star formation rate. 

The combination of feedback from supernovae ejecta with the material ejected from young massive stars in the surroundings results in the expanding bubbles around the super star clusters \citep{2004MNRAS.348..406H, 2024ApJ...973L..55L}. The evolution of galaxies depends not only on the star formation process but also on the way the various feedback effects of star formation primarily influence the interstellar medium (ISM). The overall evolution of galaxies depends on the rate at which interstellar gas in the galaxies is converted into stars, and this efficiency of star formation depends on the rate at which diffuse interstellar matter is collected into star-forming regions. 

X-ray observations expose an extended halo of soft X-ray emission around the majority of starburst galaxies \citep{2004ApJS..151..193S,2004ApJ...606..829S,2012NewA...17..524V}.  In starburst galaxies, young massive stars inject metal-enriched gas and its kinetic energy into their surrounding medium through supernovae and stellar winds \citep{1990ApJS...74..833H}. The high spatial resolution of observation facilities in the X-ray band such as \chandra and {\it XMM-Newton} provides the best facilities to study the spatial distribution of the temperature and metallicity of hot gas along these outflows \citep{2004ApJS..151..193S,2004ApJ...606..829S,2004MNRAS.349..722R,2007PASJ...59S.269T,2008MNRAS.386.1464R}. 

The second largest galaxy in the Messier~81 (M~81) group, M~82 (also known as NGC~3034 or the Cigar galaxy) is a starburst galaxy with a large number of identified supernovae. The quintessential starburst galaxy, M~82 exhibits the superwind phenomenon \citep{1978Natur.274...37A,1987AJ.....93..264M,1988Natur.334...43B,1990ApJS...74..833H,1995ApJ...439..155B,1997A&A...320..378S,1998ApJ...493..129S}. The massive ionised cloud in the halo of M~82 is the source of the X-ray emission, which is caused by shock heating \citep{1999ApJ...523..575L}. \cite{2020ApJ...904..152L} find a direct connection between the M~82 superwind and the warm-hot, metal-rich circumgalactic medium (CGM). \cite{2005ApJ...619L..99H} highlights the close morphological association between the dust and the hotter phases of the winds investigated in H$_{\alpha}$ and the X-ray emission.

This study provides the first AstroSat Ultra-Violet Imaging Telescope (UVIT)-based spatially resolved SFR measurements combined with \chandra X-ray spectroscopy for M 82. The primary objective of this study is to identify the aggregations of the star-forming regions within the M~82 galaxy using FUV observations. Moreover, we can obtain the morphological correlation between different phases of the ISM by combining FUV, H$_{\alpha}$, Mid-Infra Red (MIR) and X-ray observations. Lastly, we also determine the metal abundances from archival \chandra X-ray data of the M 82 galaxy.  

This paper is structured as follows: Section~\ref{sec2} describes the data analysis techniques. In Section~\ref{sec3}, results and and its implications are presented. Lastly, Section~\ref{sec4} encapsulates the main findings.

For M~82 we adopt a more accurate luminosity distance of 3.53 Mpc from \citep{2005AJ....129..178K}, rather than the distance listed in NASA/IPAC Extragalactic Database (NED)\footnote{\url{https://ned.ipac.caltech.edu/}} for assumed cosmology, along with a unit conversion scale of 1\arcsec\, = 17 pc.

\section{Observations and Data Reduction}
\label{sec2}
\indent

M~82 is one of the nearest starburst galaxies at a redshift $z$ = 0.000897 \citep{2008A&A...477L...5V}. We observed M~82 (proposal ID : A04\_222) galaxy using the Indian multi-wavelength astronomical satellite AstroSat \citep{2006AdSpR..38.2989A}. Observations were carried out using UVIT onboard AstroSat along with the optical filter FUV (F148W). The observations were conducted on June 01, 2018, for a total exposure time of 388 seconds. The UVIT instrument has an impressive resolution of $\sim$1.2\arcsec\, for NUV and $\sim$1.4\arcsec\, for FUV channels. The UVIT consists of two telescopes that are co-aligned with diameters of 38 cm, one is dedicated to FUV observations and the other is used for observations in the NUV and visible channels. The photometric calibrations and details of UVIT instrumentation are found in \cite{2017CSci..113..583T, 2017JApA...38...28T, 2017AJ....154..128T}.

\begin{figure*}[tbp]
  \centering
  \includegraphics[scale=0.42]{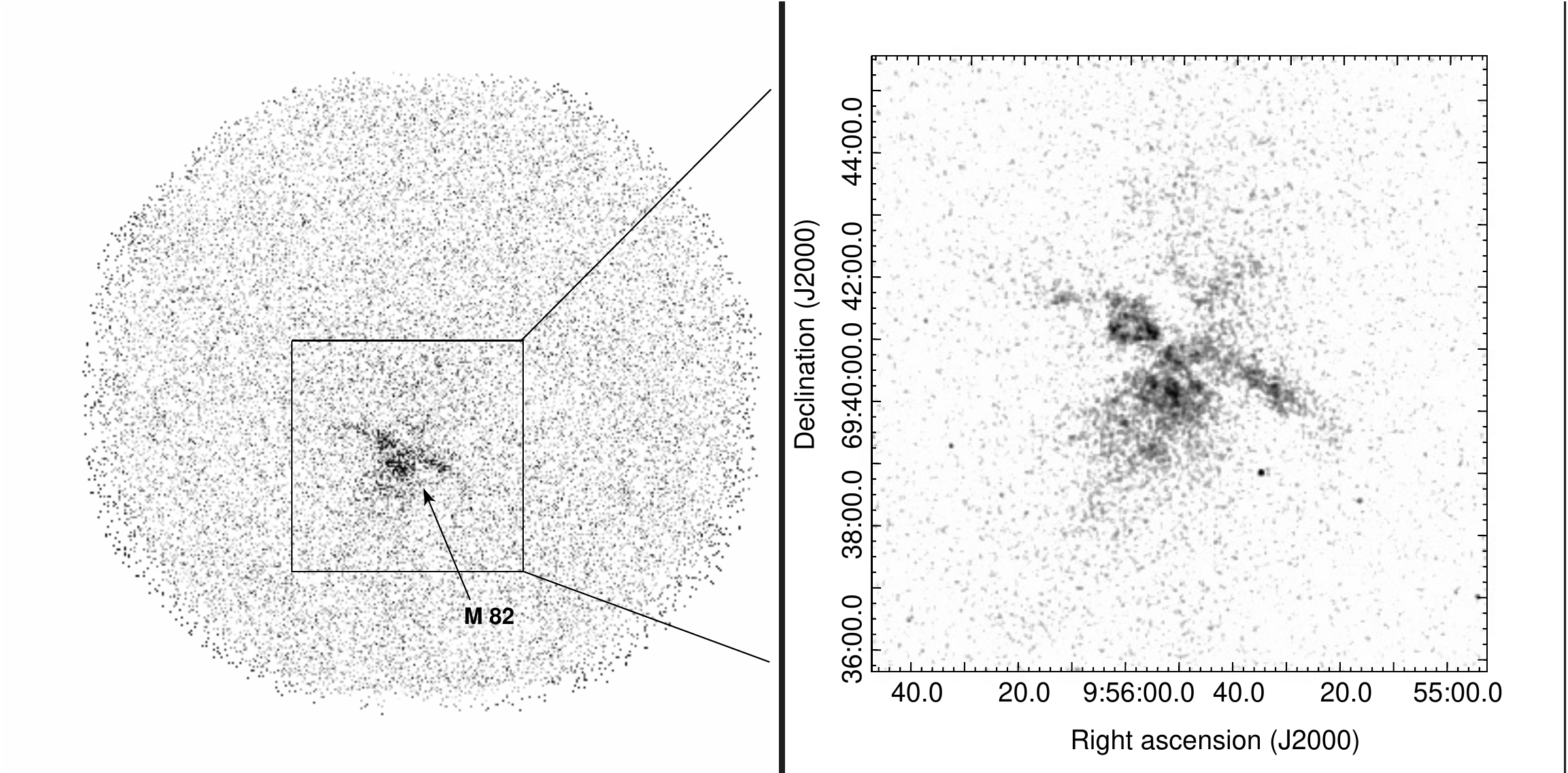}
  \caption{AstroSat$/$UVIT images of the M~82 galaxy. The left panel image shows 28 arcmin diameter field of view and the right panel shows the zoomed in view covering a $10 \times 10$~arcmin$^{2}$ region.}
  \label{fig:raw}
\end{figure*}

The observed data was analysed and reduced by following the standard routine described in \cite{2017JApA...38...28T, 2017AJ....154..128T}. The level-2 file was then obtained from the level-1 data by applying the UVIT level-2 Pipeline (UL2P) task. This thread consists of primary reduction, masking the bad pixels, flagging multi-photon event blocks, detecting cosmic-ray events, and finally correcting the drift to the event centroids. Then, multiple orbital images were combined after the astrometric correction. The final image was obtained in the form of an intensity map in units of counts per second. This paper uses only FUV(148W) band data sets, while other FUV data were rejected due to poor signal. Although we proposed the NUV for observation, the NUV observation was not possible due to observational limitations.  

This paper also uses the archival X-ray data taken from \chandra space telescope on 24 June 2009 for 120 ks exposure (ObsID : 10542) and 1 July 2009 for 120 ks (ObsID : 10543). Both observations were carried out in a very faint mode (VFAINT) on the ACIS-S chip. The X-ray data were analysed using the standard method given by \chandra X-ray observatory analysis guide, and reprocessed by using Chandra Interactive Analysis of Observations (CIAO-4.11) software package. The analysis of the X-ray data followed the procedures outlined in \cite{2019MNRAS.485.1981V, 2021SerAJ.202...17K}. Individual observational data were reprocessed from level-1 event files using the \texttt{chandra$\_$repro} task and created the level-2 file. The level-2 file was then used to identify and remove the flares in the light curve if present in the observational data by using \texttt{deflare} script provided by CIAO. To improve the signal-to-noise level, the cleaned data of both the observations were merged by using \texttt{merge$\_$obs} task. In the present study, we are interested only in the nature of the hot gas distribution, hence we identify and remove the the resolved point sources detected by using the wavelet detection algorithm \texttt{wavdetect} task. The holes left behind after removal of point sources were filled by using the \texttt{dmfilth} task. Further these cleaned image is used for imaging and spectroscopic purpose.

\section{Results and Discussion}
\label{sec3}
\subsection{Star formation in M~82}

The left panel of Figure~\ref{fig:raw} shows the 28 arcmin diameter field of view image centred on the M~82 galaxy whereas the right panel of the figure shows a zoomed-in $10 \times 10$~arcmin$^{2}$ image of the galaxy, which was lightly Gaussian smoothed to suppress noise for better visualisation. The emission features of the warm FUV emitting gas in the M~82 galaxy is dominated by a biconical structure that defines the superwinds (i.e., looks like a butterfly). This indicates that the gas distribution is primarily along the minor axis (for more details, see Section~\ref{MISM}).

The young ($<$ 1 Gyr) massive hot star in galaxies emits most of its light in the UV region. Hence, the measurement of UV flux is a reasonable estimation of current/ongoing SFRs. This study uses FUV flux to estimate the star formation rate \citep{2006ApJS..164...38I,2008MNRAS.390.1282C} from the star formation region in the M~82 galaxy, assuming a constant SFR over the past $10^8$ years.

\begin{equation}
\mathrm{SFR}_{\mathrm{FUV}}\, [M_{\odot}\,\mathrm{yr}^{-1}] = \frac{L_{\mathrm{FUV}}\, [\mathrm{erg}\,\mathrm{s}^{-1}]}{3.83 \times 10^{33}} \times 10^{-9.51}
\end{equation}

\begin{figure}[ht]
  \centering
  \includegraphics[scale=0.6]{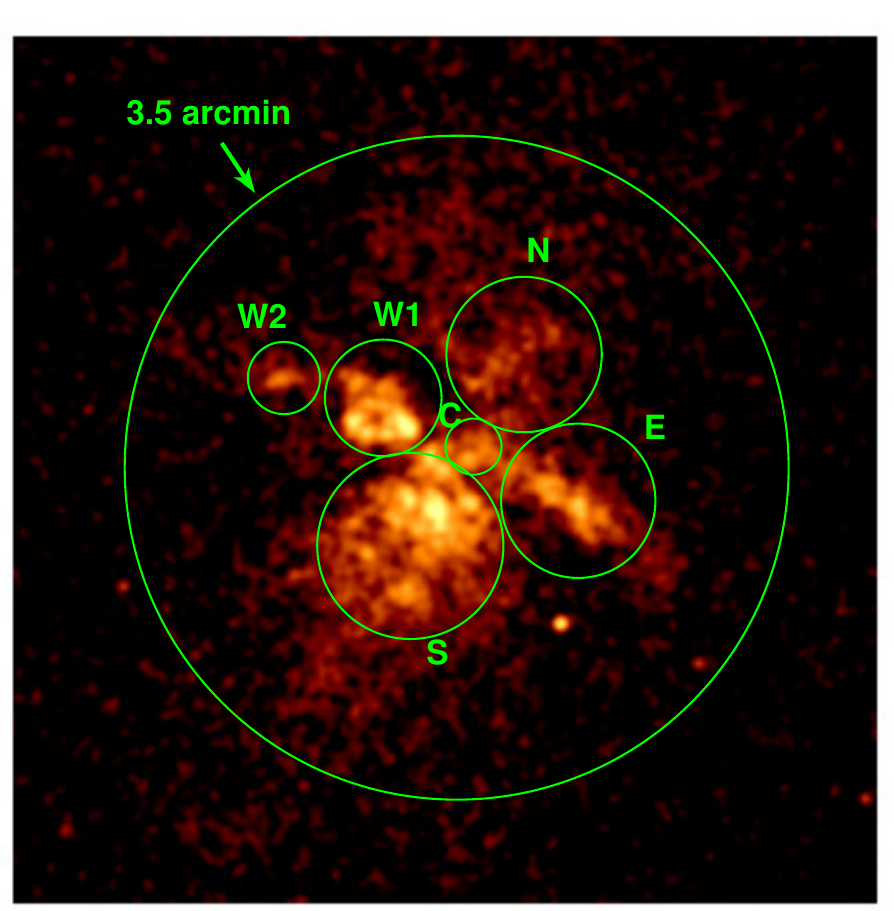}
  \caption{The figure shows the FUV Gaussian smoothed image of M 82, with the regions where flux were estimated indicated using green circles.}
  \label{fig:reg}
\end{figure}

The net background subtracted counts (net count rate) were estimated by selecting the source and background regions independently. The source regions were selected by a visual inspection method as shown in Figure~\ref{fig:reg}. The background counts were extracted from the source free regions which are far from the source and have similar region size to that of source. This count rate was further converted to flux density and flux using the unit conversion in the unit of erg cm$^{-2}$ s$^{-1}$ \AA$^{-1}$ as suggested in \cite{2017JApA...38...28T}. The obtained flux in the interested region was then converted to luminosity and finally to star formation rate. Figure~\ref{fig:reg} shows the regions of interest to estimate the SFR, and Table~\ref{tab:srf} reports the details of the estimated flux, luminosity, and SFRs in the selected regions. The total SFR in the M~82 galaxy was estimated by selecting the region up to where the FUV diffuse emission is visible (see Figure~\ref{fig:reg}). The estimated value of total SFR for the radii of 3.6 kpc (3.5\arcmin) from the centre is 0.022 M$_{\odot}$ yr$^{-1}$.

\begin{sidewaystable*}
\small	
\centering
\caption{Luminosity and star formation rates estimated from the clumpy regions of the M~82 galaxy.}
\begin{tabular}{ccccccccc}
\hline
Regions & srcs-cnts & bkg-cnts & net-cnts & cnts$\times$CF & Flux ($F$) &
$L = F \times 4\pi R^2$ & SFR \\
 & (counts) & (counts) & (counts) & ($10^{-15}$) &
($10^{-11}$ erg s$^{-1}$ cm$^{-2}$) &
($10^{40}$ erg s$^{-1}$) &
(M$_{\odot}$ yr$^{-1}$) \\
\hline
Centre (C)   & 1.33  & 0.30 & 1.03 & 3.18 & 0.470 & 0.70 & 0.0006 \\
North (N)    & 5.16  & 1.93 & 3.22 & 9.96 & 1.47  & 2.19 & 0.0018 \\
South (S)    & 14.41 & 3.03 & 11.38 & 35.2 & 5.21 & 7.76 & 0.0063 \\
East (E)     & 5.80  & 2.05 & 3.75 & 11.6 & 1.71 &  2.55 & 0.0021 \\
West-1 (W1)  & 5.18  & 1.34 & 3.83 & 11.8 & 1.75 &  2.61 & 0.0021 \\
West-2 (W2)  & 0.41  & 0.09 & 0.32 & 0.99 & 0.147 & 0.22 & 0.0002 \\
Total        & 77.83 & 38.13 & 39.70 & 123.0 & 18.2 & 27.12 & 0.0220 \\
\hline
\end{tabular}

\vskip0.2cm
\begin{minipage}{0.85\linewidth}
\small \textit{Notes:} srcs-cnts = source counts; bkg-cnts = background counts;
net-cnts = background-subtracted source counts; CF = conversion factor from
counts to flux units ($3.09 \times 10^{-15}$); Flux = cnts$\times$CF$\times\lambda$.
\end{minipage}

\label{tab:srf}
\end{sidewaystable*}

The young OB star clusters are surrounded by hydrogen gas. These young stars are hot enough to produce significant fluxes of ionising radiation. The HII region characterises the line emission; that is, the zone of ionised gas that emits the H$_{\alpha}$. This allows us to directly convert H$_{\alpha}$ line emission into the star formation rate. Thus, the photo-ionisation rate can be expressed as hydrogen recombination line flux, and this relates to the SFR by assuming an electron temperature T$_{e}$ = 10$^4$ K and by

\begin{equation}
\frac{\mathrm{SFR}_{H_{\alpha}}}{M_{\odot}\,\mathrm{yr}^{-1}} = 5.37 \times 10^{-42} \left( \frac{L_{H_{\alpha}}}{\mathrm{erg}\,\mathrm{s}^{-1}} \right)
\end{equation}

\begin{figure}
  \centering
  \includegraphics[scale=0.70]{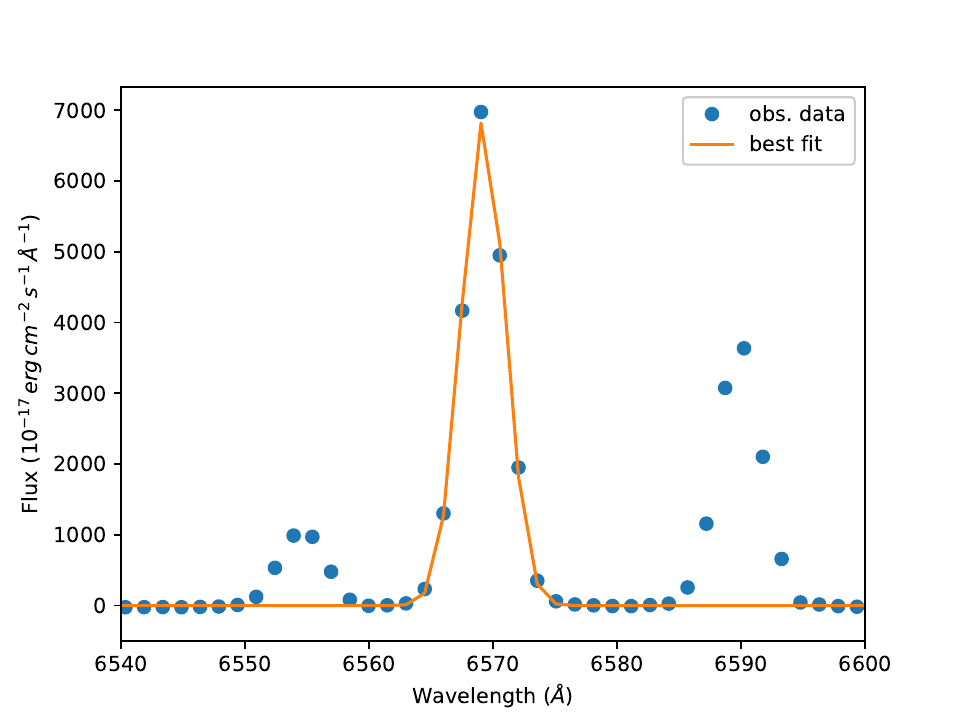}
  \caption{SDSS continuum subtracted H$_{\alpha}$ emission fitted with Gaussian 1D model.}
  \label{fig:Ha}
\end{figure}

The relation above shows that the SFR is proportional to the H$_{\alpha}$ luminosity. The line emission flux is found using the SDSS spectra (Plate-1879; MJD-54478; FiberID-411). The SDSS-I/II fibre have a 3 arcsec diameter which covers 51 pc for M~82 galaxy. The H$_{\alpha}$ line emission flux emitted from core of M~82 galaxy is estimated by fitting a single Gaussian. The best-fit model parameters are amplitude = 2.99$\pm$0.04$\times$10$^{-13}$ erg cm$^{-2}$ s$^{-1}$ \AA$^{-1}$, central wavelength = 6569.19 $\pm$ 0.03 $\AA$ and equivalent width = 1.75 $\pm$ 0.02 $\AA$. The best-fit model is shown in Figure~\ref{fig:Ha}. From this, we derive the H flux $_{\alpha}$ and find it to be 1.31 $\times$ 10$^{-12}$ erg cm$^{-2}$ s$^{-1}$ for this SDSS fiber aperture. This corresponds to luminosity L$_{H_{\alpha}}$ = $1.95\times\,10^{39}$ erg s$^{-1}$ and SFR = 0.010 M$_{\odot}$ yr$^{-1}$. \cite{1999ApJ...523..575L} derived the total H$_{\alpha}$ flux for the entire region of M~82 as $\approx$ 4.6 $\times$ 10$^{-11}$ erg cm$^{-2}$ s$^{-1}$ hence the L$_{H_{\alpha} tot}$ = $6.85\times\,10^{40}$ erg s$^{-1}$ this gives the total SFR $\sim$0.37 M$_{\odot}$ yr$^{-1}$.

The infrared (IR) emission process of galaxies is dominated by different physical mechanisms. In particular, the mid-IR (5-25 $\mu$m) emissions are dominated by the warm dust emission process, which originates from the small size of dust grains which are heated by high energetic photons generated by young massive stars. Therefore, it is safe to assume that the IR emission is mostly powered by star formation \citep{1998ARA&A..36..189K}. However, recent studies show that the far-IR emission originates from star formation activities \citep{2009ApJ...692..556R,2010A&A...518L..33H,2012ApJ...759..139K,2013MNRAS.434.2426F}. \cite{2009ApJ...692..556R} estimate the SFR by measuring emission at 24$\mu$m that results from increasing the emission by warm dust heated by the young massive stars, over the cold dust heated by the radiation field. Further, we assume that the UV emission contribution to the total luminosity powered by young star is about 20$\%$. Hence, the relation between SFR$_{\rm FIR}$ and FIR luminosity is given by \cite{2009ApJ...692..556R}

\begin{equation}
    \mathrm{SFR}_{\mathrm{FIR}}\,(M_{\odot}\,\mathrm{yr}^{-1}) = (1.12 - 1.23) \times 10^{-9} \, L(24\,\mu\mathrm{m},\, L_{\odot})
\end{equation}

The $L(24\mu \mathrm{m})$ luminosity is in units of L$_{\odot}$ and derived from the $24\mu \mathrm{m}$ flux, which was obtained from AKARI observations using the infrared camera (IRC) \citep{2010A&A...514A..14K}, and is reported in the NED database. The size of the aperture region (d) to estimate the flux is taken to be less than 4 arcmin. The $24 \mu \mathrm{m}$ flux of M~82 within the aperture region is $4.74 \times 10^{-8} \, \mathrm{erg \, s^{-1} \, cm^{-2}}$, which corresponds to a luminosity of $1.46 \times 10^{10}$ L$_{\odot}$, providing an estimate of the SFR$_{\rm FIR}$ of $\sim$ $16-18$ M$_{\odot}$ yr$^{-1}$.

A significant difference is observed between the SFRs derived from FUV, H${\alpha}$, and FIR emission. Also, the similar difference in values are obtained by \cite{2009ApJ...703.1672K} for sample of nearby galaxies. For M~82 the FUV and H${\alpha}$ based estimates of 0.022~M$_\odot$~yr$^{-1}$ and 0.010~M$_\odot$~yr$^{-1}$, respectively, are considerably lower than the FIR-derived value of 16–18~M$_\odot$~yr$^{-1}$. This discrepancy arises because the FUV and H${\alpha}$ tracers are highly sensitive to unobscured star formation and are significantly affected by dust attenuation, whereas the FIR emission primarily originates from dust heated by young massive stars, thereby tracing the dust-obscured component of star formation. Since M~82 is a prototypical dusty starburst galaxy, a substantial fraction of ongoing star formation is hidden from UV and optical observations but well captured in FIR observations. Additionally, no extinction corrections have been applied to the FUV and H${\alpha}$ estimates in this study, which further contributes to their lower SFR values. Similar trends have been reported in previous studies (e.g., \cite{1998ARA&A..36..189K}; \cite{2009ApJ...692..556R}), emphasising the need for multi-wavelength measurements to accurately quantify the total star formation rate in dusty starburst galaxies like M~82.

\subsection{X-ray Spectroscopy}
The protostarburst galaxy M~82, with its powerful superwind (Galactic-scale outflows), is a perfect target to study the driving plasma of superwinds \citep{2014MNRAS.437L..76L}. Superwinds are important feedback processes in galaxy evolution that are driven by stellar winds from massive stars and core-collapse supernovae (SN) from active star-forming galaxies \citep{1995ApJ...448...98H,2005ARA&A..43..769V,2014MNRAS.437L..76L}. The superwinds will eject metals, which will enrich the halo gasses and intergalactic medium \citep{2011Sci...334..948T}. Hence, one of the primary indicators of the starburst phenomena is the X-ray emission.

Background subtracted, point source removed source spectra were used in \texttt{XSPEC} version: 12.9.0 \citep{1996ApJ...462L..75A} for spectral fitting. A circular region of 3.5 arcmin radius centred on the source position was used for spectral extraction. The multi-temperature variable abundance, along with the power-law model, was used to fit the extracted spectra simultaneously. We employed this model because both observational and simulation studies typically shows that the central region of starburst galaxies experience a multi-temperature gas phase, spanning from cold dust to hot outflows \citep{2015ApJ...814...83L,2025ApJ...982...28L}.

\begin{figure}
\centering
\includegraphics[width=110mm,height=90mm]{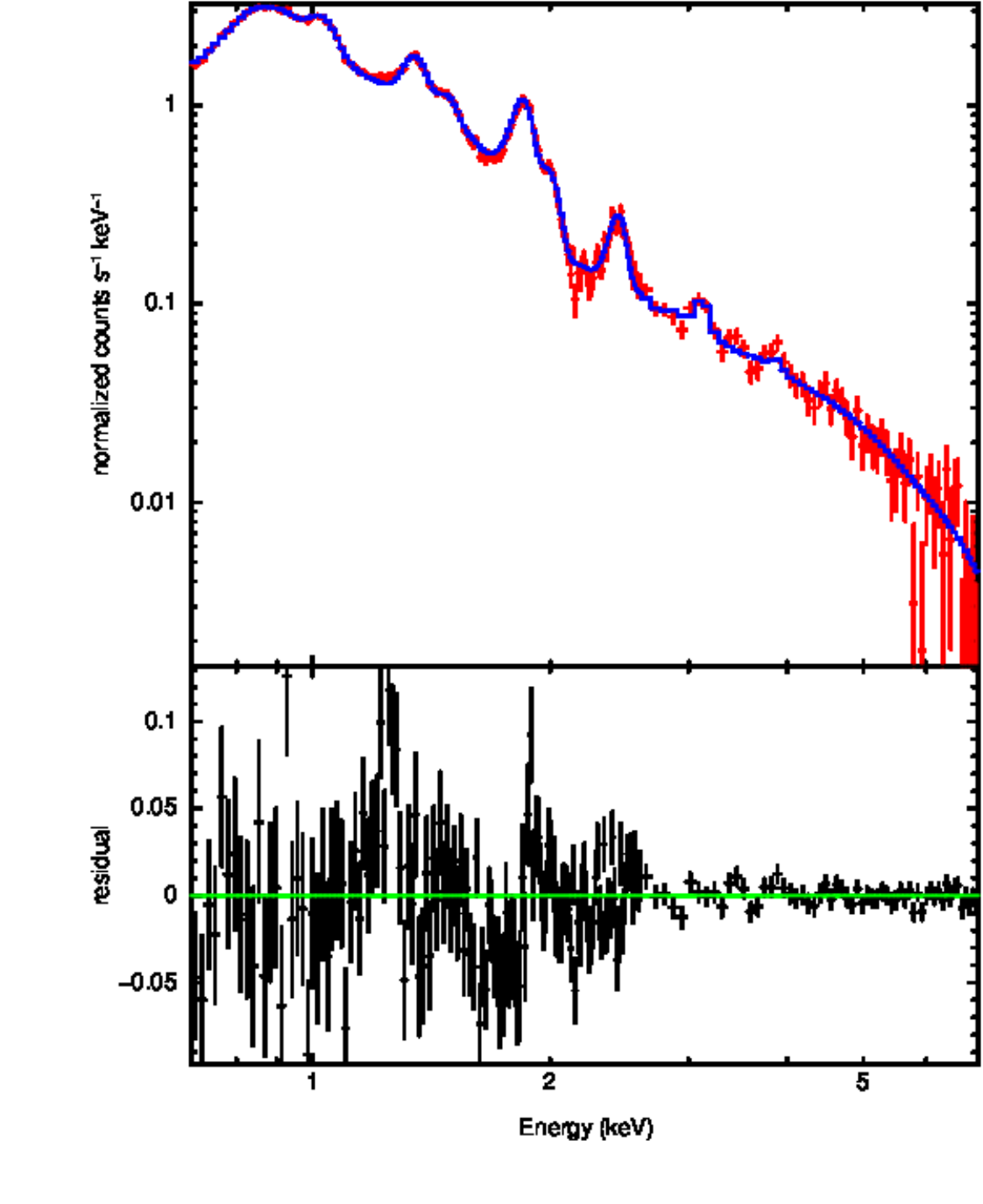}
\caption{\chandra X-ray spectrum of M~82 fitted with the best-fit model shown in blue. Bottom panel shows residual of the fit, illustrating the goodness of fit.}
  \label{fig:xspec}
\end{figure}

A detailed X-ray spectroscopic study of the M~82 galaxy is presented in \cite{2020ApJ...904..152L}. This model also includes two absorption components (\texttt{PHABS}): one is to account for the Galactic absorption along the direction of M~82, which was kept fixed at N$_H$ = 4.0 $\times$ 10$^{20}$ cm$^{-2}$ \citep{1990ARA&A..28..215D} and the other is for the intrinsic absorption which was allowed to vary. The variable collisional ionised plasma (\texttt{VAPEC}) model was used to estimate the temperature and metal abundance within the selected region. The power-law photon index was fixed at 1.5 to account for the unresolved sources present in that region. Thus the model used for spectral fitting of M~82: \texttt{phabs $\times$ phabs $\times$ (vapec + vapec + vapec + powerlaw)}. During the spectral fitting, all the temperatures kT$_1$, kT$_2$, and kT$_3$ and metal abundance Ne, Mg, Si, and S are allowed to vary, and other abundances were fixed at 0.5 $Z_\odot$. The best-fit model parameters are shown in Table~\ref{tab:spec}.  The best spectral fit using this model yielded a reduced chi-square of $\sim$ 1.35 ($\chi^{2}$ / dof = 227/168). The best-fitting spectrum is shown in Figure~\ref{fig:xspec}. The luminosity was estimated to be 2.44$\times$10$^{40}$ erg s$^{-1}$. 

We derived best-fit temperatures of kT$_1 \sim 0.22$ keV, kT$_2 \sim 0.56$ keV, and kT$_3 \sim 1.43$ keV, indicating a multiphase, mass-loaded starburst driven wind. In particular, the 0.22 keV component displays a wider spread compared to the 0.56 keV and 1.43 keV plasmas \citep{2020ApJ...904..152L}. The hottest phase (kT$_3 \sim 1.43$ keV; $T \sim 1.7 \times 10^{7}$ K) likely traces the recent supernova heated ejecta and the fast central wind material, while the intermediate temperature component (kT$_2 \sim 0.56$ keV; $T \sim 6.5 \times 10^{6}$ K) represents shock heated ISM and mass-loaded gas at the base of the outflow. The coolest component (kT$_1 \sim 0.22$ keV; $T \sim 2.6 \times 10^{6}$ K) is consistent with cooled, entrained halo gas and possible interface emission between hot and cold phases \citep{2024ApJ...968...54P}. The derived abundances indicate sub-solar neon (Ne $\sim 0.36\,Z_{\odot}$) and nearly solar to slightly super-solar $\alpha$-elements (Mg $\sim 0.86\,Z_{\odot}$, Si $\sim 0.89\,Z_{\odot}$, and S $\sim 1.23\,Z_{\odot}$), suggesting that the hot gas is predominantly enriched by core-collapse supernovae (Type-II supernovae) associated with the ongoing starburst activity \citep{2013PASJ...65...44M}. The relatively low neon abundance supports a young enrichment timescale, consistent with minimal contribution from intermediate mass stars \citep[see][]{2007PASJ...59S.269T,2024A&A...686A..96F}. These results support a scenario in which hot, $\alpha$-enhanced supernova ejecta interact with and mass-load the surrounding ISM, driving the multiphase bipolar outflow observed in M\,82.

\begin{table}
\caption{Spectral fit parameters}
\label{tab:spec}
\centering
\begin{tabular}{ll}
\hline
Parameter & Value \\
\hline
$N_{\mathrm{H}}$ ($10^{20}\,\mathrm{cm^{-2}}$) 
& $38.0^{+0.06}_{-0.06}$ \\
kT$_{1}$ (keV) 
& $0.22^{+0.05}_{-0.04}$ \\
kT$_{2}$ (keV) 
& $0.56^{+0.15}_{-0.08}$ \\
kT$_{3}$ (keV) 
& $1.43^{+0.45}_{-0.17}$ \\
Ne ($Z_{\odot}$) 
& $0.36^{+0.22}_{-0.36}$ \\
Mg ($Z_{\odot}$) 
& $0.86^{+0.06}_{-0.05}$ \\
Si ($Z_{\odot}$) 
& $0.89^{+0.93}_{-0.06}$ \\
S ($Z_{\odot}$) 
& $1.23^{+0.42}_{-0.19}$ \\
\hline
\end{tabular}

\vspace{2mm}
\parbox{0.9\linewidth}{
\footnotesize
$N_{\mathrm{H}}$ is the hydrogen column density. kT$_{1}$, kT$_{2}$, and kT$_{3}$ denote the plasma temperatures. Elemental abundances are given relative to the solar value.
}
\end{table}

\begin{figure*}
  \includegraphics[width=70mm,height=70mm]{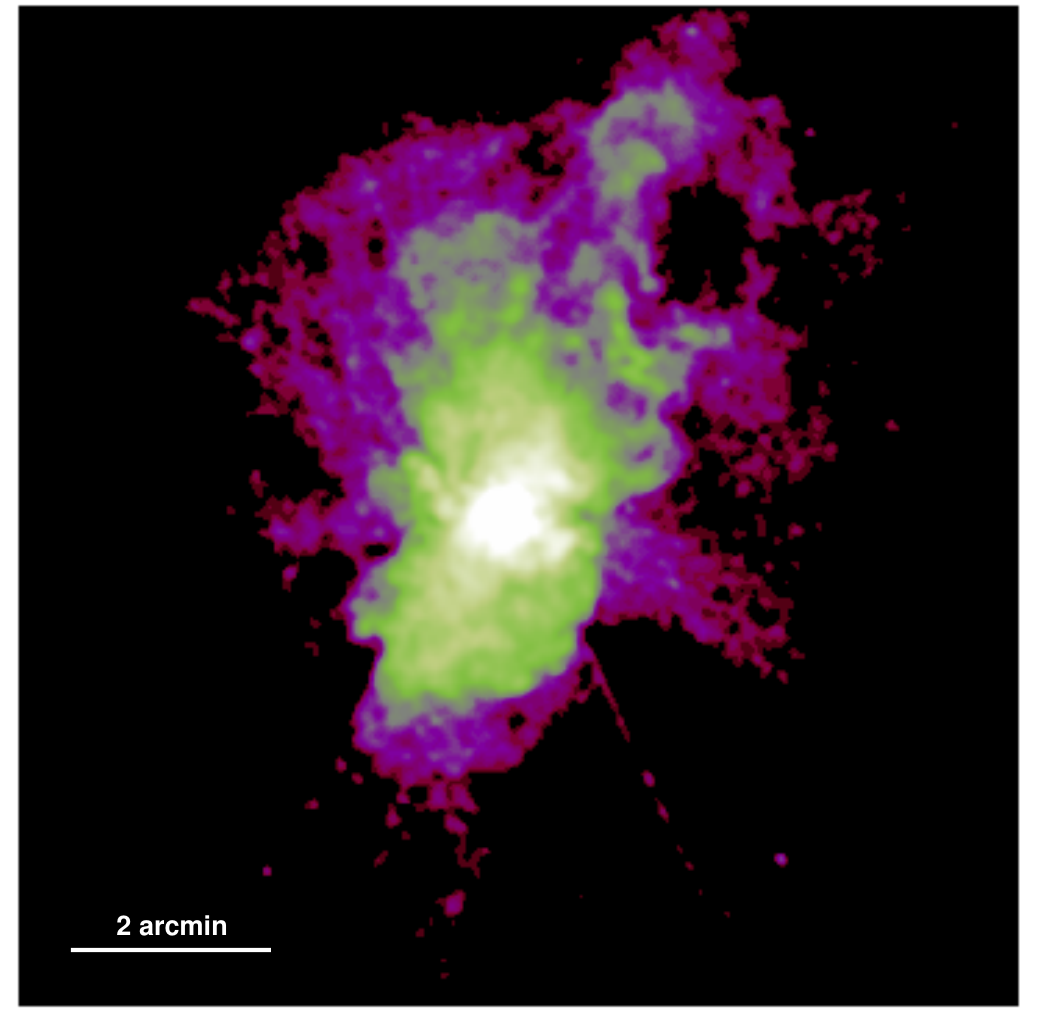}
  \includegraphics[width=70mm,height=70mm]{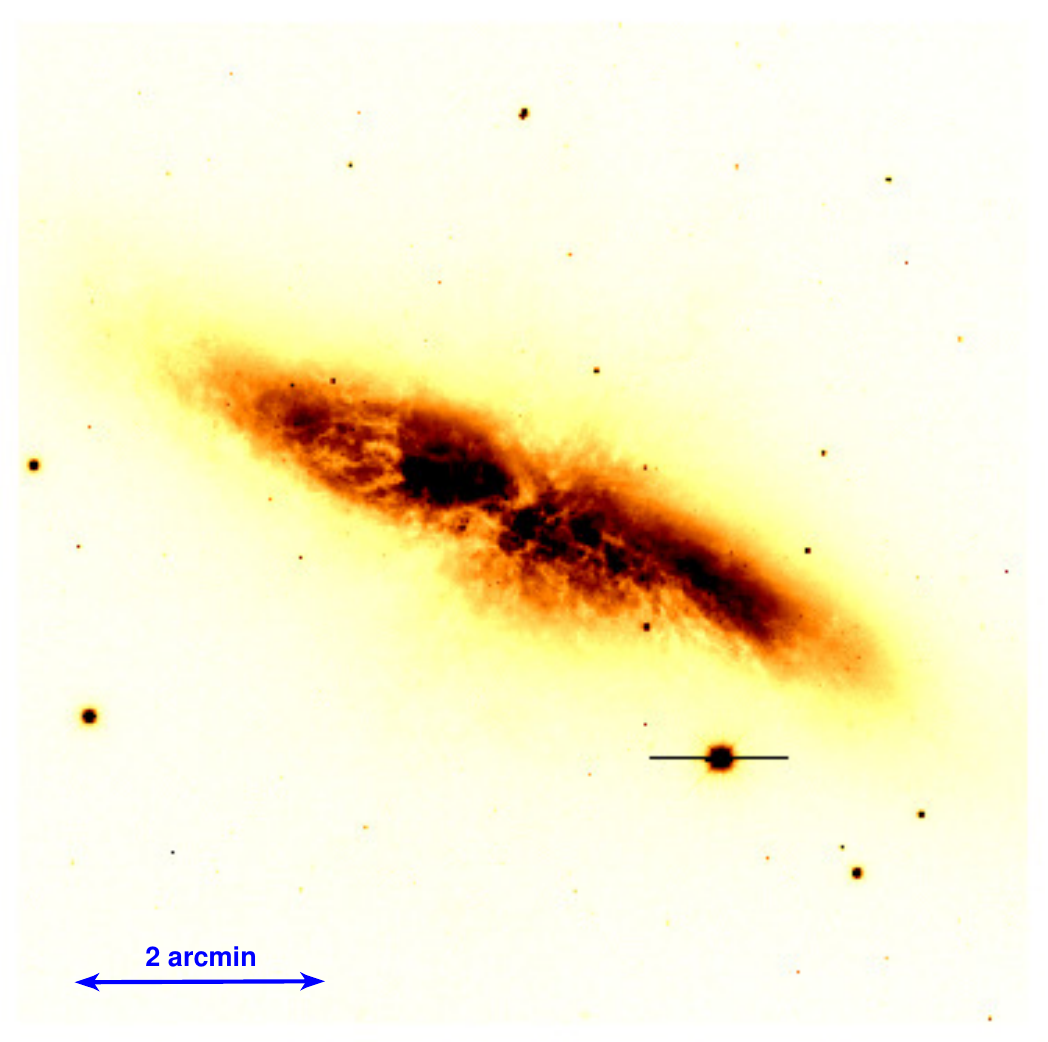}
  \caption{Left panel: \chandra 0.5 - 3 keV background subtracted, exposure corrected smoothed X-ray image. Right panel: KPNO {\it B}-band optical image.}
  \label{fig:xrayimg}
\end{figure*}

\begin{figure*}
  \centering
  \includegraphics[scale=0.50]{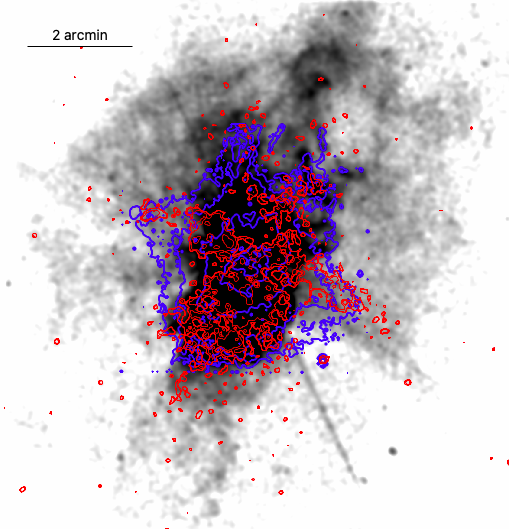}
  \caption{\chandra X-ray image in the 0.5 - 3 keV energy band, overlaid with KPNO continuum subtracted H$_{\alpha}$ contours in blue and FUV contours in red.}
  \label{fig:HaXfuv}
\end{figure*}

\subsection{Multiphase ISM}
\label{MISM}

M~82 is a galaxy that is currently experiencing an active starburst with significant outflow of material along its minor axis \citep{2015ApJ...814...83L,2020ApJ...904..152L,2025ApJ...989..100L}. We can use the multi-wavelength observational imaging data to examine the association of ISM in the M~82 galaxy with other phases. For this, we derive the hot gas emission map, which is point sources subtracted, 0.5 - 3.0 keV Gaussian smoothed \chandra ACIS image of the M~82 galaxy, shown in the left panel of Figure~\ref{fig:xrayimg}. It is evident that the distribution of hot X-ray emitting gas is complex, abundant in the nuclear region, and significantly extended along the minor axis (North-South distance $\sim$8.0 kpc) of the galaxy. Along the north side of the galaxy, the hot gas distribution displays a filamentary structure. The right panel of Figure~\ref{fig:xrayimg} shows the Kitt Peak National Observatory (KPNO) {\it B}-band image extended along NW to SE direction shows the stellar emission is obscured by the interstellar dust. Figure~\ref{fig:HaXfuv} shows the 0.5 - 3 keV X-ray image, overlaid with continuum-subtracted H$_{\alpha}$ contours from the KPNO 2.1m archival observations in blue and FUV contours in red. Figure~\ref{fig:HaXfuv} illustrates a notable spatial correlation between the warm gas, traced by FUV and H$\alpha$ emissions, and the hot gas, traced by soft X-ray emissions. This suggests that the multiphase ISM in M~82 is physically interconnected through a starburst-driven outflow. The emission from X-ray, FUV, and H$_{\alpha}$ are all extended in the minor axis direction. The alignment of these emissions along the minor axis, combined with significant dust obscuration along the major axis, supports the scenario of a bipolar superwind driven by starburst activity, which transports mass, energy, and metals from the central region to the halo and CGM.

\section{Conclusion}
\label{sec4}

In this work, we performed a multi-wavelength investigation of the starburst galaxy M 82 using UVIT FUV imaging and \chandra X-ray data, including both imaging and spectroscopy, to study its star-forming regions, hot gas distribution, and feedback driven outflows.

\begin{itemize}

\item The spatially-resolved FUV analysis identifies several star-forming knots, yielding a total SFR of 0.022 M$_{\odot}$ yr$^{-1}$ within $\sim$3.6 kpc.

\item Comparison with H$_{\alpha}$ (0.010 M$_{\odot}$ yr$^{-1}$) and IR-derived (16 - 18 M$_{\odot}$ yr$^{-1}$) SFRs highlights that a significant fraction of star formation is hidden by dust.

\item FUV, H$_{\alpha}$, and soft X-ray emission exhibit a strong morphological correlation along the minor axis, tracing multi-phase outflows driven by intense star formation.

\item X-ray spectral modelling indicates enhanced $\alpha$-element abundances (Ne, Mg, Si, S), consistent with enrichment by Type-II supernovae in the starburst core.

\end{itemize}

These findings provide new insights into the multi-phase ISM and the role of stellar feedback in shaping the evolution of starburst galaxies. Future deep UVIT observations combined with integral field spectroscopy and high-resolution IR mapping will further clarify the interaction between star formation, outflows, and metal enrichment in M~82.

\bmhead{Acknowledgements}
The authors are grateful to the anonymous referee for their careful reading and insightful comments. This research has made use of data obtained from AstroSat-UVIT observations, \chandra Archival observations, SDSS, and KPNO data.  We also thank \chandra software provided by the \chandra X-ray Center (CXC) in the application packages CIAO and Sherpa. This paper also uses information from NASA extragalactic data centre (NED). The research work done at the Physical Research Laboratory, Ahmedabad, is funded by the Department of Space, Government of India.  NV and MKP thanks to IUCAA, Pune, for providing the library facility.









\def\nat{Nature}%
\def\aj{AJ}%
\def\actaa{Acta Astron.}
\def\araa{ARA\&A}
\def\apj{ApJ}
\def\apjl{ApJ}
\def\apjs{ApJS}
\def\aap{A\&A}
\def\aapr{A\&A~Rev.}
\def\aaps{A\&AS}
\def\apss{Ap\&SS}
\def\baas{BAAS}
\def\caa{Chinese Astron. Astrophys.}
\def\cjaa{Chinese J. Astron. Astrophys.}
\def\icarus{Icarus}
\def\jcap{J. Cosmology Astropart. Phys.}
\def\jrasc{JRASC}
\def\memras{MmRAS}
\def\mnras{MNRAS}
\def\na{New A}
\def\nar{New A Rev.}
\def\pra{Phys.~Rev.~A}
\def\prb{Phys.~Rev.~B}
\def\prc{Phys.~Rev.~C}
\def\prd{Phys.~Rev.~D}
\def\pre{Phys.~Rev.~E}
\def\prl{Phys.~Rev.~Lett.}
\def\pasa{PASA}
\def\pasp{PASP}
\def\pasj{PASJ}%
\def\sovast{SOVAST}%
\def\ssr{Space Sci. Rev.}
\def\physrep{Phys. Rep.}

\bibliography{sn-bibliography}

\end{document}